\documentclass[11pt,a4paper]{article} 
\usepackage{ol}
\usepackage{graphicx}
\usepackage{amsmath}
\newcommand{\micron}{\,\mathrm{\mu m}}

\begin{document}

\title{Electro-optically tunable microring resonators in lithium niobate} 

\author{Andrea Guarino$^{*}$, Gorazd Poberaj, Daniele Rezzonico, Riccardo Degl'Innocenti and Peter Gunter}

\address{Nonlinear Optics Laboratory, Institute of Quantum Electronics, \\ETH Zurich, 8093 Zurich, Switzerland\\ \texttt{guarino@phys.ethz.ch}}


\vspace{0.5 cm}
\noindent\textbf{Optical microresonators have recently attracted a growing attention in the photonics community\cite{vahalareview}. Their applications range from quantum electro-dynamics to sensors and filtering devices for optical telecommunication systems, where they are likely to become an essential building block\cite{Erice}. The integration of nonlinear and electro-optical properties in the resonators represents a very stimulating challenge, as it would incorporate new and more advanced functionality. Lithium niobate is an excellent candidate material, being an established choice for electro-optic and nonlinear optical applications. Here we report on the first realization of optical microring resonators in submicrometric thin films of lithium niobate. The high index contrast films are produced by an improved crystal ion slicing and bonding technique using benzocyclobutene. The rings have radius $\mathbf{R=100}\,\mathbf{\mu m}$ and their transmission spectrum has been tuned using the electro-optic effect. These results open new perspectives for the use of lithium niobate in chip-scale integrated optical devices and nonlinear optical microcavities.}

The established use of wavelength division multiplexed (WDM) for local area network systems has raised the demand for new filtering and switching functions\cite{WDM}. In order to integrate these devices on a wafer scale, whispering gallery mode microresonators represent the most compact and efficient solution. They consist of a bus waveguide evanescently coupled to a micrometer-size ring resonator; the characteristic size-dependent frequency spectrum of the ring allows only selected wavelength channels to be transmitted or shifted to another waveguide.  Small radii allow a large free spectral range - i.e. large separation between the filtered channels - but increase the propagation bending losses\cite{Marcuse}, which can compromise the quality factor Q - i.e. the wavelength selectivity - of the device. To overcome this limitation, high refractive index contrast between the ring core and the surrounding materials is mandatory. A second, very important, requirement relates to the tunability. The possibility to electrically control the transmission spectrum, via electro-optic effect, would allow extremely compact and ultrafast modulation and switching. By integrating arrays of microring resonators on a single optical chip, the realization of complex functions would be feasible\cite{resonatorarray}. Besides, large-Q resonators based on non centro-symmetric materials would exploit the high amount of stored energy for enhancing the efficiency of nonlinear optical phenomena\cite{nonlinearrings}.

\begin{figure*}[hpt]
\centering
\includegraphics[width=13cm]{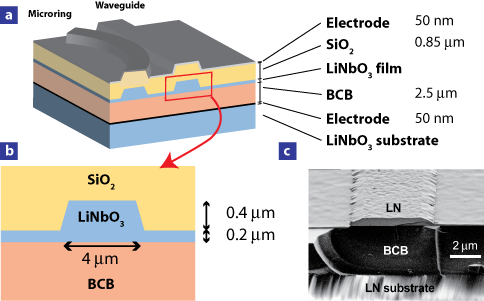}
\caption{\textbf{Cross-section of a lithium niobate microring resonator structure.} \textbf{a-b,} Schematic layout and cross section of a microring resonator and waveguide. The waveguide and ring core consist of structured lithium niobate thin film, bonded using BCB polymer to a lithium niobate wafer and covered by SiO$_{2}$. The upper and lower electrodes enable the application of an electric field along the z-axis of lithium niobate. \textbf{c,} Scanning electron microscopy image, viewed at an angle, of a cleaved end before the deposition of the oxide and upper electrode layers.}
\label{structure}
\end{figure*}

Several examples of microring resonators have been proposed and successfully realized in the last years in a variety of materials like semiconductors\cite{siliconringnature,silicononinsulator,siliconring,siliconcascaded}, silica\cite{silica} and polymers\cite{driessen,polymer}.The advanced structuring technology in semiconductor materials enables the realization of very high-Q resonators even for radii as small as $10\,\mathrm{\mu m}$. Silicon-based resonators can be tuned by electrically-driven carriers injection in the core\cite{siliconmz}, but do not own truly nonlinear optical properties and their application is limited to infrared wavelengths. Polymers represent a very flexible solution in terms of processing and structuring, but the minimum resonator dimensions (and therefore the maximum achievable free spectral range) are limited by the low refractive index of the material. Silica rings, finally, do not provide any fast nonlinear or electro-optical property.

We propose lithium niobate as a very attractive new choice for microresonating devices. It has the potential for ultrafast modulation since it has large electro-optic coefficients\cite{LNbook,mojcaLN} ($r_{33}=31\,\mathrm{pm/V}$, $r_{13}=8\,\mathrm{pm/V}$ ), large transparency range ($0.4-5\,\mathrm{\mu m}$) and a wide intrinsic bandwidth. It is well known in existing electro-optic and nonlinear optical applications, and large dimension wafers of crystalline quality are available. A new technique, based on crystal ion slicing and wafer bonding, has been recently developed to produce sub-micrometric thin films of single-crystalline quality\cite{slicing,payammethod}; it provides much higher refractive index contrast than the standard waveguide production methods in lithium niobate. This is an essential asset for the fabrication of small radius ring resonators, as we show in the Supplementary Information, Section 1.  An electro-optic modulator has been demonstrated\cite{payammodulator} by using lithium niobate films bonded to SiO$_2$ as substrate. However, the direct bonding method does not provide large area films and lacks of sufficient reproducibility, due to the severe requirements on the surface roughness and imperfections. 
Bonding of lithium niobate films to other substrates (for instance, semiconductors) has also been  reported\cite{lnvahala}, but suffers of film cracking due to the large mismatch between the thermal expansion coefficients of films and substrates and does not provide the optical contrast needed for the realization of optical microresonators.

\begin{figure*}[t]
\centering
\includegraphics[width=13cm]{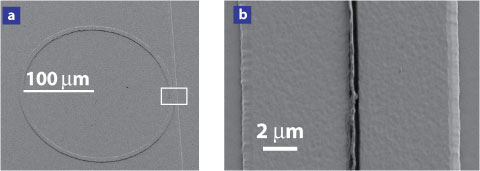}
\caption{\textbf{Structured lithium niobate microring resonator.} Scanning electron microscopy images of (\textbf{a}) lithium niobate microring resonator with radius $R=100\micron$ and (\textbf{b}) enlargement of the coupling region between the waveguide and the ring. The gap size is approximately $0.2\micron$. }
\label{ringandgap}
\end{figure*}

We have improved the lithium niobate thin film fabrication technique by introducing the use of benzocyclobutene (BCB), a well known adhesive polymer\cite{BCB1,BCB2} for the realization of 3D semiconductor devices, to successfully and reproducibly bond large area  ($>1.5\,\mathrm{cm^2}$) submicrometric films. Full details of the fabrication procedure are presented in the Methods section. The films are realized by implanting z-cut lithium niobate wafers with He$^+$ ions which accumulate below the surface. The ion energy ($E=195\,\mathrm{keV}$ in our experiments) determines the position of their density peak (here, $0.68\micron$). Subsequently, a sample of the implanted wafer is cut and bonded to another lithium niobate wafer, covered by a metallic electrode and a BCB layer (approximately $2.5\micron$). The bonded pair is thermally treated for several hours; this step is crucial: on one hand it strengthens the bonding by curing the polymer, on the other it causes helium bubbles to aggregate and leads to splitting of the film. Finally, it also provides partial annealing of the defects introduced by ion implantation. The use of BCB offers several advantages: its planarization and adhesion properties reduce the role of surface defects and greatly enhance the reproducibility and the size of the transferred films; optically, BCB has excellent transparency in the visible and infrared region, and as a substrate provides a suitable optical confinement due to its low refractive index ($n\approx1.55$).

After the splitting, the film thickness is reduced by Ar$^{+}$ ion-etching of a sacrificial layer of approximately $60\,\mathrm{nm}$. This step reduces the surface roughness inherently induced by the straggling of the implanted ions. 
The waveguides and the rings are structured by photo-lithographic techniques explained below; the ridge height is $0.4\,\mathrm{\mu m}$, as a compromise between a low surface scattering from the lateral walls and the need for a suitable lateral confinement. Finally, a $0.85\,\mathrm{\mu m}$-thick covering SiO$_2$ layer reduces the scattering losses and ensures optical insulation between the core and the upper chromium electrode. We emphasize that the geometry chosen allows the applied electric field to be along the $z$-axis of lithium niobate and therefore to exploit the electro-optic coefficient $r_{33}=31\,\mathrm{pm/V}$. 

A schematic representation of this device can be seen in Fig.\ref{structure}a-b. Fig.\ref{structure}c is a scanning electron microscopy (SEM) image of one end face of the structure cleaved before the deposition of the oxide and upper electrode layers.

\begin{figure*}[htp]
\centering
\includegraphics[width=14cm]{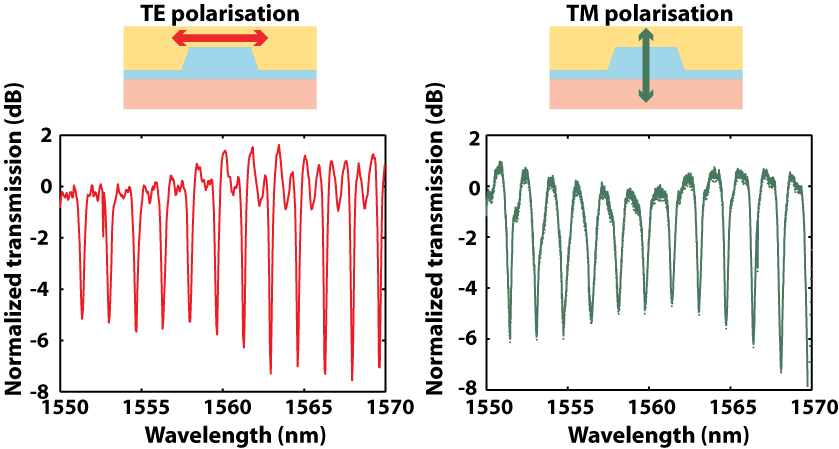}
\caption{\textbf{Transmission spectrum of a $100\micron$-radius ring resonator.} The measured normalized transmitted light at the through port for both TE (left) and TM (right) modes using a tunable source in the $\lambda=1.55-1.57\micron$ region is shown. The free spectral range is $1.66\,\mathrm{nm}$ and the finesse $5$. The modulation depth is approximately $7\,\mathrm{dB}$.}
\label{transmission}
\end{figure*}

The high-refractive index contrast structures produced with this technique ($\Delta n\approx0.65$) are ideally suited for the realization of microresonators: the numerical calculations presented in the Supplementary Information, Section 1, show that the bending losses are negligible even for ring radii of $10\,\mu m$. The high contrast also implies stringent conditions on the waveguide dimensions to obtain single-mode operation (see Supplementary Information, Section 2 for details). The submicrometric thickness of our films support only one guided mode in the vertical direction. Single-mode operation in the horizontal direction requires a waveguide width of approximately $w\approx1\micron$, which is too narrow for standard lithographic techniques. Our waveguides have a width of approximately $w=4\,\mathrm{\mu m}$, hence they are multi-mode. However, the results demonstrate that in our structures the contribution of higher-order modes is nearly negligible,  because these modes have higher propagation losses. More sophisticated structuring techniques (laser or electron-beam lithography) could potentially achieve true single mode operation without excessive scattering losses.

Another critical issue in the structuring of microresonators relates to the coupling coefficient between the waveguide and the resonator. To maximise the light extinction at the resonant wavelength, the coupling should be equal to the total propagation loss per resonator round trip. The horizontal coupling geometry requires a very accurate separation between the ring and the waveguide. To achieve a sub-micrometer gap, we lithographically define the waveguides and the rings in two steps, using a negative-tone photoresist. In the first step the straight waveguides are created  in the photoresist using mask photolithography and hardening. Subsequently, the rings are formed on a second photoresist  layer with the same procedure and positioned using a standard mask-aligner. The two-step technique, similar to the one presented in a recent work\cite{Daniele}, reduces the diffraction effects that would inhibit the formation of the narrow gap if a single-step illumination was used. The structures are then transferred into lithium niobate by Ar$^+$ ion etching. The scanning electron micrographs of Fig.\,\ref{ringandgap} show a structured micro-ring resonator in lithium niobate with radius $R=100\micron$ (a) and a sub-micrometer gap (b) obtained by this technique.


The measured transmission spectrum of a coupled ring resonator around $\lambda=1.55\,\mathrm{\mu m}$ is presented in Fig.\,\ref{transmission}. Both TE (electric field direction mainly parallel to the film) and TM (perpendicular to the film) polarisations of the waveguide bus can be coupled into the cavity and show the distinctive features of a microresonator. The extinction ratio at the resonant wavelengths is approximately $7\,\mathrm{dB}$. The free spectral range of the resonator is about $\Delta\lambda_{FSR}\approx 1.66\,\mathrm{nm}$, as predicted by the calculations presented in the Supplementary Information, Section 3, which account for the modal dispersion of the structure. The resonator finesse is approximately $F=\Delta\lambda_{FSR}/\delta\lambda_{FWHM}\approx5$ and the corresponding Q value is $Q=4\times10^3$. This value is probably limited by implantation-induced defects and scattering losses. The propagation losses were measured to be approximately $4\,\mathrm{cm^{-1}}$.

\begin{figure}[ptb]
\centering
\includegraphics[width=6.5 cm]{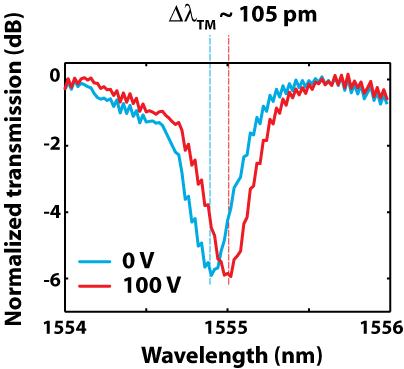}
\caption{\textbf{Electro-optic shift of the resonance curve.} Resonance curve at a wavelength around $1.555\micron$ (blue) and the corresponding electro-optically shifted curve (red) by applying a voltage $V=100\,\mathrm{V}$ to the device electrodes. The shift corresponds to an approximate tunability of $0.14\,\mathrm{GHz/V}$}
\label{figshift}
\end{figure}

The electro-optic properties of lithium niobate microrings have been tested by shifting the transmission spectrum applying a static electric field to the device electrodes. In this preliminary configuration the electrodes are placed over the whole device surface below the polymer cladding layer and above the SiO$_{2}$ buffer layer, respectively. In the Supplementary Information, Section 4, we show in detail how the induced refractive index change affects the resonance condition for both TE and TM modes. The resonance of a TM mode displayed in Fig.\,\ref{figshift} shows a  $\Delta\lambda=105\,\mathrm{pm}$-shift in response to an applied voltage of $\Delta V=100\,\mathrm{V}$. This wavelength shift corresponds to frequency tunability of $0.14\,\mathrm{GHz/V}$. This value indicates a reduction of the electro-optic activity of our structure by approximately 50\% compared to the bulk material. A partial decrease of the electro and nonlinear optical properties in lithium niobate thin films due to implantation-induced defects has already been reported in a previous work\cite{slicingmodulator,annealing}. A complete restore of the electro-optic coefficient is possible with a post-slicing annealing of the film at high temperature\cite{slicingmodulator} (800$^{\circ}$C), which is however higher than the maximum temperature allowed by our current adhesive polymer ($320^{\circ}$C).

We propose the following strategy to reduce the switching voltage for a specific wavelength channel. First, an optimization of the polymer and oxide thickness is required to increase the electric field in the lithium niobate film, which is smaller than in the underlying polymer due to the large lithium niobate dielectric constant\cite{mojcaLN} ($\epsilon_{33}=28$).  More specifically, numerical simulations show that the polymer thickness can be safely reduced to $0.8\,\mathrm{\mu m}$ with a negligible effect on the propagation losses. The upper electrode can be replaced by a semitransparent electrode directly in contact with the lithium niobate layer. These steps, and an optimized annealing process would lead to a tunability larger than $1\,\mathrm{GHz/V}$. Second, the required wavelength shift can be reduced with advanced lithographic techniques by increasing the Q-factor of the cavity. A Q-factor of $2\times10^{4}$ and an enhanced tunability of $1\,\textrm{GHz/V}$ would decrease the switching voltage below $10\,\mathrm{V}$ still preserving a theoretical maximum modulation speed of $10\,\mathrm{Gbit/s}$. Indeed, such high Q-factor would rise the requirements on the laser wavelength control. Nevertheless, LiNbO$_{3}$-based microring resonators may become a primary choice for the realisation of highly-compact electro-optic filtering devices for local area network applications, since they have a largely reduced size than existing MZI-based modulators. 

Beside the potential application for local area WDM systems, our first demonstration of an electro-optical microresonators in lithium niobate may open new perspectives for integrated nonlinear photonic devices. Microcavities based on non-centrosymmetric materials may enable ultra-compact second-harmonic generation and optical parametric generation. The latter has already been demonstrated in a LiNbO$_{3}$ resonator\cite{maleki}, yet in a device whose size is 15 times larger than our microring resonators.

\noindent\textbf{Acknowledgments} We are grateful to the AIM team at the Research Center Rossendorf, Germany, for performing the He$^{+}$ implantation of LiNbO$_{3}$ wafers in the frame of the RITA Program, Contract No. 025646. We also thank S. Reidt for depositions of the electrodes, J. Hajfler for professional polishing of the samples and C. Herzog, M. Jazbinsek and L. Mutter for helpful discussions. This work was supported by ETH Research Grant TH-13/05-2.

\subsection*{Methods}
Here we describe in detail the device fabrication.
The implanted wafer is a pure congruent lithium niobate z-cut wafer (Crystal Technology, Inc.). The He$^{+}$ ions had energy $E=195\,\mathrm{keV}$. The implantation fluence was $\phi=4\times10^{16}\,\mathrm{ions/cm^2}$ and the sample holder was heated to $T=100^{\circ}\,\mathrm{C}$ during the process. The implanted wafer was cut in $12\times14\,\mathrm{mm^{2}}$ pieces and cleaned using standard RCA1 solution. 
The substrate consists of another pure congruent z-cut lithium niobate wafer. The bottom electrode was formed by deposition of a $50\,\mathrm{nm}$-thick chromium layer.
BCB, under its commercial name of Cyclotene 3022-46 (Dow Chemical) was spun at 4000 rpm, after the use of the Adhesion Promoter AP3000. The polymer thickness was approximately $2.5\,\mathrm{\mu m}$.

Thermal treating of the bonded pair was performed at $T=290^\circ\mathrm{C}$ in vacuum conditions (to avoid BCB oxidation) for several hours. No bonding pressure was applied during this step. The splitted film was subsequently smoothed by sputtering of Ar$^+$ ions for $50$ minutes ($200\,\mathrm{W}$), which removed approximately $60\,\mathrm{nm}$ of material. Atomic force microscope measurements demonstrate that the RMS surface roughness was reduced by 40\% to $4\,\mathrm{nm}$ by this process.

The photolithographic structuring of waveguides and rings, the negative tone photoresist SU-8 was used in two steps. In each step, the photoresist layer was $1.4\,\mathrm{\mu m}$-thick and after illumination and development the structures are hard baked at $120^\circ\,\mathrm{C}$. The positioning of the samples was performed using a Karl-Suss MJB3UV300 mask-aligner.

The ridges and rings were transferred into lithium niobate after $320$ minutes of $200\,\mathrm{W}$ etching using Ar$^+$ ions. After removing the remaining SU-8, the sample was covered by a PECVD-layer of SiO$_2$ of approximately $0.85\,\mathrm{\mu m}$. The upper electrode was deposited with the same parameters as the bottom electrode.

Finally the sample was sawed and the sides were polished to ensure efficient end-fire coupling. Typical sample length is $3\,\mathrm{mm}$.

\subsection*{Experiments}
The microrings were tested using a tunable laser diode Santech TSL-220. The tuning range is $1.530-1.610\,\mathrm{\mu m}$ and the spectral width is $1\,\mathrm{MHz}$. The light was spatially filtered using a single mode fiber and end-fire coupled into the waveguide using a 100x microscope objective with NA=0.9. The transmitted light was collected using a 40x (NA=0.45) long working-distance microscope objective.

\newpage

\appendix
\section*{Supplementary Information}
\section{Bending losses and Q-factors of high-index contrast resonators}
In this section we show how a high refractive index contrast between the lithium niobate and the surrounding material minimise the bending losses and therefore can lead to an improved Q-factor of these microring resonators.

\begin{figure}[h!]
\centering
\includegraphics[width=12cm]{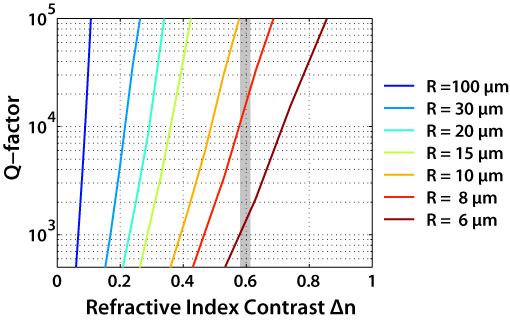}
\caption{Q-factor of a microring resonator (due to bending losses only) as a function of the refractive index contrast $\Delta n=n_{core}-n_{substrate}$ for different ring radii. The calculations refer to a TM mode at $\lambda=1.55\,\mathrm{\mu m}$.  The core dimensions are $w=1\,\mathrm{\mu m}$ and $h=0.6\,\mathrm{\mu m}$ for $\Delta n=0.65$ and are scaled to be single mode for every value of $\Delta n$.  The grey line shows that for lithium niobate microring resonators, surrounded by BCB polymer, small ring radii are possible because $\Delta n\approx 0.6$. }
\label{bendlosses}
\end{figure}

Decreasing the microring radius is attracting, because the free spectral range (FSR), defined as the separation between two adjacent resonant wavelengths, increases and may become even larger than the wavelength range used for WDM applications. For example, a lithium niobate microring resonator with radius $R=10\,\mathrm{\mu m}$ will have a FSR of approximatively $16\,\mathrm{nm}$. However, one of the most critical factor limiting the minimum useful ring radius are the bending losses.  These can be qualitatively understood by describing the bend as a straight waveguide, where the effective index is a decreasing function in the radial direction\cite{Marcuse}. This implies that at a certain distance from the waveguide core, the solution of the Maxwell equations becomes a radiating field; this radiation is a loss source, as in a leaky waveguide. The larger the refractive index difference, the smaller will be the leakage. 

The losses essentially determine the Q-factor of the optical cavity: if only propagation losses $\alpha$ are present, the Q factor can be calculated using\cite{Erice}:
\begin{equation}
Q\approx\frac{\pi N}{\alpha \lambda}
\end{equation}
where $N$ is the effective index and $\lambda$ is the vacuum wavelength.
A bending loss of $\alpha=10\,\mathrm{dB/cm}$ would already imply that $Q\approx4000$ without any other loss source.

Using a commercial software\cite{Selene}, we have calculated the bending losses for a single mode waveguide as a function of the refractive index difference between the core and the surrounding material, assuming a TE guided mode and a light wavelength of $\lambda=1.55\,\mathrm{\mu m}$. The results are shown in Figure \ref{bendlosses}. It is evident that lithium niobate thin films, bonded using BCB polymer, have a large advantage over other nonlinear or electro-optic materials like polymers, whose typical contrast is about $0.1-0.2$. The refractive index contrast of $\Delta n\approx0.6$ enables the fabrication of rings having a radius as small as $R=10\,\mathrm{\mu m}$, if other loss sources can be neglected.

\section{Conditions for single mode operation in lithium niobate waveguides}
The high refractive index difference between the lithium niobate ($n_{o}=2.21$ and $n_{e}=2.13$ at $\lambda=1.55\,\mathrm{\mu m}$)\cite{LNbook} and the BCB polymer ($n\approx1.55$) requires the film thickness and the waveguide width to be smaller than a limiting value (about $1\,\mathrm{\mu m}$) to achieve single-mode operation. 
\begin{figure}[h!]
\centering
\includegraphics[width=13cm]{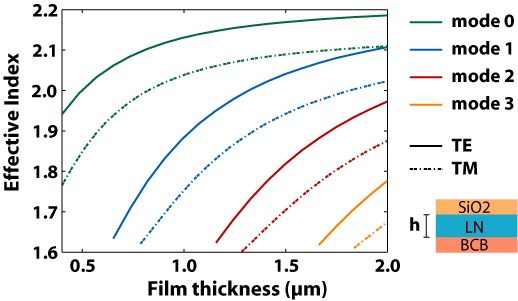}
\caption{Effective index of the first guided TE (straight line) and TM (dashed line) modes in a lithium niobate waveguide as a function of the film thickness. The waveguide is surrounded by SiO$_{2}$ (top) and bonded using BCB (bottom); the modes are calculated at $\lambda=1.55\,\mathrm{\mu m}$. In case of a thickness $h=0.6\,\mathrm{\mu m}$, the confinement of any higher order mode is either not existing or very weak. } 
\label{modesheight}
\end{figure}

\begin{figure}[h!]
\centering
\includegraphics[width=13cm]{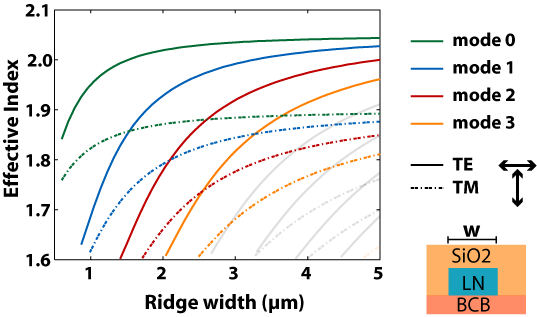}
\caption{Effective index of the first guided TE (straight line) and TM (dashed line) modes in a lithium niobate waveguide as a function of the width $w$ for a film thickness $h=0.6\,\mathrm{\mu m}$. The waveguide is surrounded by SiO$_{2}$ (top) and bonded using BCB (bottom); the modes are calculated at $\lambda=1.55\,\mathrm{\mu m}$. Single mode operation can be obtained only with structures narrower than $1\,\mathrm{\mu m}$.}
\label{modeswidth}
\end{figure}

We have calculated the modal curve at $\lambda=1.55\,\mathrm{\mu m}$ for planar lithium niobate waveguides surrounded by SiO$_{2}$ and bonded using BCB. The effective index dependence on the film thickness is presented in Figure \ref{modesheight}. The maximum film thickness to achieve single mode operation is given by $h=0.7\,\mathrm{\mu m}$. With the technique explained in the paper, the film thickness is determined by the energy of the implanted ions. Our films, being created using $E=195\,\mathrm{keV}$ and subsequently Ar$^{+}$ sputtered, have a thickness $h=0.6\,\mathrm{\mu m}$ and therefore support only one guided mode. 

The dependence of the effective index on the waveguide width, assuming a waveguide height $h=0.6\,\mathrm{\mu m}$ is shown in Figure \ref{modeswidth}. As it can be seen, a single-mode operation requires waveguide structures narrower than $1\,\mathrm{\mu m}$.This condition is not met in our microrings, which have $w\approx4\,\mathrm{\mu m}$ and therefore support several modes. However, single-mode operation could be reached by using more advanced structuring techniques, like laser or electron beam lithography. Nevertheless, in our experiments higher order modes have limited impact, since they have higher propagation losses.

\section{Group effective index and calculation of the free spectral range}
To properly analyse the transmission properties of the microring resonators in thin films of lithium niobate, the dispersion properties of the guided modes have to be considered. We calculate in this section the modal dispersion of our structure, i.e. the sensitivity of the effective index of the guided modes to wavelength changes. 

The resonance condition of the microring resonator is given by:
\begin{equation}
\label{res}
\frac{2 \pi N(\lambda) L}{\lambda}=2\pi m \quad \mathrm{or} \quad
N(\lambda) L = m \lambda
 \end{equation}
where $N$ is the effective index of the guided mode, $L$ is the resonator's length, $\lambda$ is the light wavelength and $m$ is an integer number.

The effective index depends on the wavelength by two distinct mechanisms. First, the refractive index of the materials are wavelength dependent (material dispersion); second, the guiding properties of the structure (i.e. the solutions of the propagation equation) depend on the wavelength (modal dispersion). The second mechanism is specially relevant, due to the tiny vertical dimension of the core (the film thickness) with respect to the wavelength.

The free spectral range (FSR) is defined as the difference between two adjacent resonant wavelengths $\Delta\lambda_{FSR}=\lambda_{m}-\lambda_{m+1}$; it can be approximately calculated by differentiating Equation (\ref{res}):

\begin{eqnarray}
\nonumber -\frac{N(\lambda_{m})}{\lambda_{m}}+\frac{N(\lambda_{m+1})}{\lambda_{m+1}}=\frac{1}{L} \\
\label{FSR}
\Delta\lambda_{FSR}\approx\frac{\lambda^{2}}{L (N-\lambda\frac{\partial N}{\partial \lambda})}=\frac{\lambda^{2}}{L N_{g}}
\end{eqnarray}
where we have defined $N_{g}:=N-\lambda\frac{\partial N}{\partial \lambda}$ as the \emph{group effective index}. This quantity contains the material and modal dispersion.

\begin{figure}[ht]
\centering
\includegraphics[width=12cm]{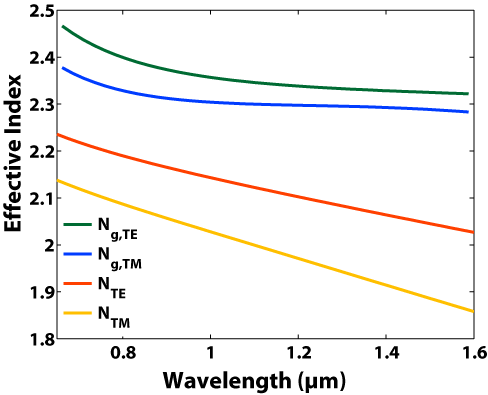}
\caption{Effective ($N$) and group effective ($N_{g}$) index as a function of the wavelength for the first TE and TM guided mode in a lithium niobate microring resonator with radius $R=100\,\mathrm{\mu m}$. Waveguide width and height are $w=4\,\mathrm{\mu m}$ and  $h=0.6\,\mathrm{\mu m}$, respectively. The waveguide core is surrounded by BCB polymer ($n\approx 1.55$) and SiO$_{2}$ ($n\approx1.45$) as explained in the article. The discrepancy between $N$ and $N_{g}$ increases with higher wavelength due to reduced guiding effect.}
\label{groupindex}
\end{figure}

We have calculated $N_{g}$ for the first guided mode of both TE and TM polarisations in the straight waveguide and in the microring resonator, considering both modal and material dispersion, using a mode solver software\cite{Selene}. The results are shown in Figure \ref{groupindex}.

The difference between the effective index $N$ and the group effective index $N_{g}$ is large (more than $25\%$), therefore the role of dispersion in lithium niobate flms shall not be overlooked for a proper calculation of the free spectral range. In the visible region of the spectrum the mode is confined and the main contribution to dispersion is given by the material dispersion. In the infrared region of interest in the paper, the main contribution is given by the modal dispersion, since the film thickness is smaller than the wavelength.

To estimate the free spectral range around $\lambda=1.55\,\mathrm{\mu m}$ we therefore use $N_{g,TE}\approx2.323$ and $N_{g,TM}\approx2.286$, which in (\ref{FSR}) yield a free spectral range of $1.65\,\mathrm{nm}$ and $1.67\,\mathrm{nm}$, respectively. This is well in agreement with the spacing experimentally measured ($1.66\,\mathrm{nm}$) as presented in the paper.

\section{Electro-optic effect in lithium niobate microresonators}
In this section we present more details about the electro-optic tuning effect induced in lithium niobate microresonators.

Applying an electric field between the upper and lower electrode of the structure shown in Figure 1 of the main text, the refractive indices of lithium niobate are changed according to its electro-optic tensor $r_{ijk}$\cite{Yariv}. This change modifies the effective index $N$ of the microring resonator, hence its round-trip phase and its resonance condition. 

Since the applied electric field is directed along the z-axis of lithium niobate, the only relevant electro-optic coefficients are $r_{13}$ and $r_{33}$, which are responsible for the change of the ordinary and extraordinary refractive index, respectively\cite{LNbook}. For a small electric field $\delta E$, the effective index $N$ varies according to:
\begin{equation}
\centering
\begin{aligned}
\delta N&=\frac{\partial N}{\partial n_{o}}\delta n_{o}+\frac{\partial N}{\partial n_{e}}\delta n_{e}=\\
&=-\bigg(\frac{\partial N}{\partial n_{o}}\frac{r_{13} n_{o}^{3}}{2}+\frac{\partial N}{\partial n_{e}}\frac{r_{33} n_{e}^{3}}{2}\bigg)\delta E
\end{aligned}
\end{equation}

The derivatives of the effective index $N$ with respect to the ordinary and extraordinary refractive indices can be determined by using a developed mode solver routine in MATLAB which accounts for the anisotropy of the structure. For the given dimensions ($h=0.6\,\mathrm{\mu m}$, $w=4\,\mathrm{\mu m}$) and the wavelength $\lambda=1.55\,\mathrm{\mu m}$ we calculated the following values result for the first guided TE and TM mode:
\begin{equation}
\begin{array}{cc}
\mathrm{TE}\left\{
\begin{aligned}
\frac{\partial N}{\partial n_{o}}&=0.97\\
\frac{\partial N}{\partial n_{e}}&=0
\end{aligned}
\right. &

\mathrm{TM}\left\{
\begin{aligned}
\frac{\partial N}{\partial n_{o}}&=0.10\\
\frac{\partial N}{\partial n_{e}}&=0.77
\end{aligned}
\right.

\end{array}
\end{equation}
We remark that the TE mode is sensitive to the ordinary refractive index only, while the TM mode is sensitive to both. The values indicate that, due to the small core dimensions, the change of the effective index is smaller than the change of the material refractive index. 

To estimate $\delta N$, the electric field in the lithium niobate core must be determined. Due to the large difference between the dielectric constant of the film ($\epsilon_{33}=28$\cite{mojcaLN}) and of the other materials ($\epsilon_{BCB}=2.6$ and $\epsilon_{SiO_{2}}=3.8$) the field in the film is considerably smaller than in the adjacent layers. The field in the i-th layer can be calculated from the applied voltage $\Delta V$ using the continuity of the vertical component of the electric displacement field $D_{i}$:
\begin{equation}
\left\{
\begin{aligned}
\sum_{i}E_{i}d_{i}&=\Delta V \\ 
D_{i}=\epsilon_{0}\epsilon_{i}E_{i}&=\mathrm{const}
\end{aligned}
\right.
\end{equation}
Therefore, the field in the lithium niobate layer is given by:
\begin{equation}
E_{LN}=\frac{\Delta V}{\epsilon_{LN}\sum_{i}\frac{d_{i}}{\epsilon_{i}}}=\frac{\Delta V}{d_{eff}}
\end{equation}
The quantity $d_{eff}$ is an effective thickness. It is a practical quantity which accounts for every layer thickness and dielectric constant if the field in the lithium niobate layer has to be determined. The effective thickness which corresponds to the structure presented in the text is $d_{eff}\approx 34\,\mathrm{\mu m}$. 
The effective index change can be finally expressed as 
\begin{equation}
\label{finaldeltaN}
\delta N=-\bigg(\frac{\partial N}{\partial n_{o}}\frac{r_{13} n_{o}^{3}}{2}+\frac{\partial N}{\partial n_{e}}\frac{r_{33} n_{e}^{3}}{2}\bigg)\frac{\delta V}{d_{eff}}
\end{equation}
The shift of the resonance wavelength, due to a perturbation of the effective index, can be derived by differentiating Equation (\ref{res}) for a fixed integer value $m$. The effect of the dispersion is taken into account by using the definition of group effective index given in the previous section:
\begin{equation}
\begin{aligned}
\delta\bigg(\frac{N(\lambda) L}{\lambda}\bigg)&=0 \\
\bigg(\frac{1}{\lambda}\frac{\partial N}{\partial \lambda}-\frac{N}{\lambda^{2}}\bigg)\delta \lambda + \frac{1}{\lambda}\delta N&=0 \quad \mathrm{hence} \quad
\delta\lambda=\frac{\lambda}{N_{g}}\delta N
\end{aligned}
\end{equation}
We emphasise that the tunability of the microring resonator does not depend on its length. Substituting in the last expression the result of (\ref{finaldeltaN}) we obtain:
\begin{equation}
\label{shift}
\delta \lambda=-\frac{\lambda}{N_{g}}\bigg(\frac{\partial N}{\partial n_{o}}\frac{r_{13} n_{o}^{3}}{2}+\frac{\partial N}{\partial n_{e}}\frac{r_{33} n_{e}^{3}}{2}\bigg)\frac{\delta V}{d_{eff}}
\end{equation}
The tunability $T$ is often expressed in terms of frequency shift per voltage $\delta\nu/\delta V$:
\begin{equation}
\label{tunability}
\frac{\delta\nu}{\delta V}=\frac{c}{\lambda N_{g}}\bigg(\frac{\partial N}{\partial n_{o}}\frac{r_{13} n_{o}^{3}}{2}+\frac{\partial N}{\partial n_{e}}\frac{r_{33} n_{e}^{3}}{2}\bigg)\frac{1}{d_{eff}}
\end{equation}
Using the bulk lithium niobate values\cite{mojcaLN} for the electro-optic coefficients and the previously determined $d_{eff}$ we can calculate $T$ and the expected wavelength shift if a voltage $\delta V=100\,\mathrm{V}$ is applied. 

\begin{center}
\begin{tabular}{|c|c|c|c|}
\hline
		mode & tunability $\frac{\delta\nu}{\delta V}$ & \multicolumn{2}{c|}{Shift $\delta\lambda$ [pm] for $\delta V=100\,\mathrm{V}$} \\
		& [GHz/V] & expected & measured \\
		\hline
		\hline
		TM & 0.28 & 224 & $105\pm10$ \\
		TE & 0.10 & 80 & $30\pm8$ \\
		\hline
\end{tabular}
\end{center}
The measured values are approximately $40-50\%$ of the expected values.
To confirm the results, alternative measurements have been performed by applying a small ($10.5\,\mathrm{V}$ amplitude) oscillating voltage and detecting the oscillating light intensity with a lock-in amplifier. The wavelength has been chosen off-resonance, to ensure the maximum response. The experiments confirmed the values determined by the resonance shift.

This decrease can be attributed to the ion implantation induced crystal defects, since the samples could not be annealed at a temperature higher than $300\,\mathrm{^{\circ}C}$.

\bibliographystyle{unsrt}
\bibliography{references}

\end{document}